\documentclass[11pt]{article}         %
\usepackage{times}                               %
\usepackage{chicagor}

\newtheorem{THEOREM}{Theorem}[section]
\newenvironment{theorem}{\begin{THEOREM} \hspace{-.85em} {\bf :} }%
                        {\end{THEOREM}}
\newtheorem{LEMMA}[THEOREM]{Lemma}
\newenvironment{lemma}{\begin{LEMMA} \hspace{-.85em} {\bf :} }%
                      {\end{LEMMA}}
\newtheorem{COROLLARY}[THEOREM]{Corollary}
\newenvironment{corollary}{\begin{COROLLARY} \hspace{-.85em} {\bf :} }%
                          {\end{COROLLARY}}
\newtheorem{PROPOSITION}[THEOREM]{Proposition}
\newenvironment{proposition}{\begin{PROPOSITION} \hspace{-.85em} {\bf :} }%
                            {\end{PROPOSITION}}
\newtheorem{DEFINITION}[THEOREM]{Definition}
\newenvironment{definition}{\begin{DEFINITION} \hspace{-.85em} {\bf :} \rm}%
                            {\end{DEFINITION}}
\newtheorem{CLAIM}[THEOREM]{Claim}
\newenvironment{claim}{\begin{CLAIM} \hspace{-.85em} {\bf :} \rm}%
                            {\end{CLAIM}}
\newtheorem{EXAMPLE}[THEOREM]{Example}
\newenvironment{example}{\begin{EXAMPLE} \hspace{-.85em} {\bf :} \rm}%
                            {\end{EXAMPLE}}
\newtheorem{REMARK}[THEOREM]{Remark}
\newenvironment{remark}{\begin{REMARK} \hspace{-.85em} {\bf :} \rm}%
                            {\end{REMARK}}

\newcommand{\thm}{\begin{theorem}}
\newcommand{\lem}{\begin{lemma}}
\newcommand{\pro}{\begin{proposition}}
\newcommand{\dfn}{\begin{definition}}
\newcommand{\rem}{\begin{remark}}
\newcommand{\xam}{\begin{example}}
\newcommand{\cor}{\begin{corollary}}
\newcommand{\prf}{\noindent{\bf Proof:} }
\newcommand{\ethm}{\end{theorem}}
\newcommand{\elem}{\end{lemma}}
\newcommand{\epro}{\end{proposition}}
\newcommand{\edfn}{\bbox\end{definition}}
\newcommand{\erem}{\bbox\end{remark}}
\newcommand{\exam}{\bbox\end{example}}
\newcommand{\ecor}{\end{corollary}}
\newcommand{\eprf}{\bbox\vspace{0.1in}}
\newcommand{\beqn}{\begin{equation}}
\newcommand{\eeqn}{\end{equation}}

\newcommand{\bbox}{\vrule height7pt width4pt depth1pt}

\newcommand{\clm}{\begin{claim}}
\newcommand{\eclm}{\end{claim}}

\newcommand{\sat}{\models}

\newcommand{\rimp}{\Rightarrow}

\renewcommand{\phi}{\varphi}

\newcommand{\A}{{\cal A}}

\newcommand{\F}{{\cal F}}

\newcommand{\I}{{\cal I}}

\newcommand{\K}{{\cal K}}

\renewcommand{\P}{{\cal P}}

\newcommand{\R}{{\cal R}}

\newcommand{\ol}{\setlength{\itemsep}{0pt}\begin{enumerate}}
\newcommand{\eol}{\end{enumerate}\setlength{\itemsep}{-\parsep}}
\newcommand{\ul}{\setlength{\itemsep}{0pt}\begin{itemize}}
\newcommand{\dl}{\setlength{\itemsep}{0pt}\begin{description}}
\newcommand{\edl}{\end{description}\setlength{\itemsep}{-\parsep}}
\newcommand{\eul}{\end{itemize}\setlength{\itemsep}{-\parsep}}

\newcommand{\commentout}[1]{}

\newcommand{\bi}{\begin{itemize}}
\newcommand{\ei}{\end{itemize}}
\newcommand{\be}{\begin{enumerate}}
\newcommand{\ee}{\end{enumerate}}

\newcommand{\im}{im}
\newcommand{\kernel}{ker}
\renewcommand{\S}{{\cal S}}

\begin{document}

\title{Anonymity and Information Hiding \\ in Multiagent Systems%
\thanks{Authors were supported in part by NSF under grant 
CTC-0208535, by ONR under grants  N00014-00-1-03-41 and
N00014-01-10-511, by the DoD Multidisciplinary University Research
Initiative (MURI) program administered by the ONR under
grant N00014-01-1-0795, and by AFOSR under grant F49620-02-1-0101.
Kevin O'Neill was also supported in part by a graduate fellowship from the
National Science and Engineering Research Council of Canada.
A preliminary version of this paper appeared at the 16th IEEE Computer Security
Foundations Workshop in Pacific Grove, California.
}
}

\author{Joseph Y. Halpern \ \ \ \ \ \ \ \ \ \ \ Kevin R. O'Neill \\
Department of Computer Science\\
Cornell University\\
$\{$halpern, oneill$\}$@cs.cornell.edu\\
}

\maketitle
\thispagestyle{empty}

\begin{abstract}
We provide a framework for reasoning about information-hiding requirements
in multiagent systems and for reasoning about anonymity in particular. 
Our framework employs the modal logic of knowledge within the context of the
runs and systems framework, much in the spirit of our earlier work on
secrecy 
\cite{HalOn02}. 
We give several definitions of anonymity with respect 
to agents, actions, and observers in multiagent systems, and we relate our definitions
of anonymity to other definitions of information hiding, such as secrecy. We also
give probabilistic definitions of anonymity that are able to quantify an observer's
uncertainty about the state of the system. Finally, we relate our definitions of
anonymity to other formalizations of anonymity and information hiding, including
definitions of anonymity in the process algebra CSP and definitions of information
hiding using function views.
\end{abstract}

\section{Introduction}

The primary goal of this paper is to provide 
a formal framework for reasoning
about anonymity in multiagent systems.   
The importance of anonymity has increased over the past few years
as more communication passes over the Internet. 
Web-browsing, message-sending, and file-sharing are all important 
examples of activities that computer users would like to 
engage in, but may be reluctant to do unless they can
receive guarantees that their anonymity will be protected to some
reasonable degree. 
Systems are being built that attempt to implement 
anonymity
for various kinds of network communication
(see, for example,
\cite{goel02,hopper03,levine02,reiter98,sherwood02,syverson97}). 
It would be helpful to have a formal framework in which to reason 
about the level of anonymity that such systems provide. 

We view anonymity as an instance of a more general problem:
information hiding.
In the theory of computer security, many of the fundamental problems and much of the 
research has been concerned with the hiding of information. Cryptography, for instance,
is used
to hide the contents of a message from untrusted observers as it passes from
one party to another. Anonymity requirements are intended to ensure that the identity
of the agent who performs some action remains hidden from other
observers. Noninterference requirements 
essentially say that {\em everything} about classified or high-level 
users of a system should be hidden from low-level users.
Privacy is a catch-all term that means different things to different
people, but 
it typically involves hiding personal or private information from others.

Information-hiding properties such as these can be 
thought of as providing answers to the following set of questions: 
\ul
\item What information needs to be hidden?
\item Who does it need to be hidden from?
\item How well does it need to be hidden?
\eul
By analyzing security properties with these questions in
mind, it often 
becomes clear how different properties relate to each other. These
questions can 
also serve as a test of a definition's usefulness: an information-hiding
property 
should be able to provide clear answers to these three questions.

In an earlier paper \cite{HalOn02}, we formalized 
secrecy 
in terms of knowledge.
Our focus was on capturing what it 
means for one agent to have {\em total\/} secrecy with respect to
another,
in the sense that no information flows from the first agent to the
second.
Roughly speaking, 
a high-level user has total secrecy 
if the low-level user never knows anything about the
high-level user that he didn't initially know.  Knowledge 
provides a
natural way to express information-hiding properties---information is
hidden from $a$ if $a$ does not know about it.  Not surprisingly, our
formalization of anonymity is similar in spirit to our
formalization of secrecy.  
Our definition of secrecy says that a classified agent maintains 
secrecy with respect to an unclassified agent if the unclassified
agent doesn't learn any new fact that depends only
on the state of the classified agent.
That is, if the agent didn't know a classified
fact $\phi$ to start with, then the
agent doesn't know it at any point in the system.
Our definitions of anonymity
say that an agent performing an action maintains anonymity with 
respect to an observer if the observer never learns certain facts
having to do with whether or not the agent performed the action.

Obviously, total secrecy and anonymity are different.
It is possible
for $i$ to 
have
complete secrecy while still not having 
very strong guarantees of
anonymity, for example,
and it is possible to have anonymity without preserving secrecy.  
However, thinking carefully about
the relationship between
secrecy and anonymity suggests new and interesting
ways of thinking about anonymity.
More generally, formalizing anonymity and information hiding in terms
of knowledge is useful for capturing the intuitions that
practitioners have.

We are not the first to use knowledge and belief to
formalize notions of information hiding.  
Glasgow, MacEwen, and Panangaden \citeyear{GMP92} describe 
a logic for reasoning about security that includes both 
{\em epistemic\/}
operators (for reasoning about knowledge) and {\em deontic\/} operators
(for reasoning about permission and obligation).
They characterize some security policies in terms of 
the facts that an agent is
permitted to know.  Intuitively, everything that an agent is not
permitted to know must remain hidden.
Our approach is similar,
except that we specify the formulas that an
agent is {\em not} allowed to know, rather than the formulas she is permitted
to know. 
One advantage of accentuating the negative is that we do not need to use
deontic operators in our logic.

Epistemic logics have also been used to define information-hiding
properties, including noninterference and anonymity.
Gray and Syverson \citeyear{gray98} 
use an epistemic logic to define probabilistic noninterference, and
Syverson and Stubblebine \citeyear{syverson99} use one
to formalize definitions of anonymity. 
The thrust of our paper is quite different from these.
Gray and Syverson focus on one particular definition of information
hiding in a probabilistic setting, while Syverson 
and Stubblebine
focus on describing an axiom system that is useful for reasoning
about real-world systems, and
on how to reason about and compose parts of the system into
adversaries and honest agents.
Our focus, on the other hand, is on giving a
semantic characterization of anonymity in a framework that lends itself
well to modeling systems.

Shmatikov and Hughes \citeyear{hughes02} position their approach to
anonymity (which is discussed in more detail in Section~\ref{sec:functionviews}) as an
attempt to provide an interface between logic-based approaches,
which they claim are good for specifying the desired properties (like
anonymity), and formalisms like CSP, which they claim are good for
specifying systems.  We agree with their claim that logic-based approaches
are good for specifying properties of systems, but also claim
that, with an appropriate semantics for the logic, there is no need to
provide such
an interface.  While there are many ways of specifying systems,
many end up identifying a system with a set of runs or traces, and 
can thus be embedded in the runs and systems framework
that we use.

Definitions of anonymity using epistemic logic are {\em possibilistic}.
Certainly, if $j$ believes that any of 1000 users (including $i$)
could have performed the action that $i$ in fact performed, 
then $i$ has some degree of anonymity with respect to $j$.  However, if $j$
believes that the probability that $i$ performed the action is 
.99, the possibilistic assurance of anonymity provides little comfort.
Most previous formalizations of anonymity have not 
dealt with probability; they
typically conclude with an acknowledgment that it is
important to do so, and suggest that their formalism can
indeed handle probability.
One significant
advantage of our formalism is that it
is completely straightforward to add probability in a natural way,
using known techniques \cite{HT}.
As we show in Section \ref{sec:probanon},
this lets us formalize the (somewhat less formal)
definitions of probabilistic anonymity given
by Reiter and Rubin \citeyear{reiter98}.
In this paper, we are more concerned with defining and specifying
anonymity 
properties than with describing systems 
for achieving anonymity or with verifying anonymity
properties. We want to define what anonymity means by using syntactic 
statements that have a well-defined semantics. Our 
work
is similar in spirit to
previous papers that have given definitions of anonymity and other
similar properties, such as the proposal for terminology given by
Pfitzmann and K\"ohntopp~\citeyear{terminology} and the
information-theoretic definitions of anonymity given by
Diaz, Seys, Claessens, and Preneel~\citeyear{Diaz02}.

The rest of this paper is organized as follows.  In
Section~\ref{sec:review} we briefly review the runs and systems
formalism of \cite{FHMV} and describe
how it can be used to represent knowledge.
In Section~\ref{sec:anonymity}, we show how anonymity can be defined
using knowledge, and relate this definition to other notions of 
information hiding, particularly 
secrecy (as defined in our earlier work).
In Section~\ref{sec:probanon}, we extend the
possibilistic definition of Section~\ref{sec:anonymity} so that it can
capture probabilistic concerns.  
As others have observed \cite{hughes02,reiter98,syverson99}, there
are a number of 
ways to define anonymity. Some definitions provide very strong
guarantees of anonymity, while others are easier to verify in practice.
Rather than giving an
exhaustive list of definitions, we focus on a few representative
notions, and 
show by example that our logic is expressive enough to capture many
other notions of interest.
In Section~\ref{sec:related}, we compare our
framework to that of three other attempts to formalize anonymity, by
Schneider and Sidiropoulos \citeyear{schneider96}, Hughes and Shmatikov
\citeyear{hughes02}, and Stubblebine and Syverson \citeyear{syverson99}.
We conclude in Section~\ref{sec:conclusion}.

\section{Multiagent Systems: A Review}\label{sec:review}

In this section, we briefly review the multiagent systems framework; we
urge the reader  
to consult 
\cite{FHMV} for more details.

A {\em multiagent system\/} consists of $n$ agents, each of which is in some
{\em local state\/} at a given point in time. We assume that an agent's
local state encapsulates all the information to which the agent has
access. In the security setting, the local state of an agent might
include initial information regarding keys, the messages she has
sent and received, and perhaps the reading of a clock. The 
framework makes no assumptions about the precise nature of the local
state.

We can view the whole system as being in some {\em global state},
a tuple consisting of the local state of each agent and the
state of the environment. 
Thus, a global state has the form $(s_e, s_1,\ldots, s_n)$,
where $s_e$ is the state of the environment and $s_i$ is agent $i$'s
state,
for $i = 1, \ldots , n$. 

A {\em run\/} is a function from time to global states. Intuitively,
a run is a complete description of what happens over time in one
possible execution of the system. A {\em point\/} is a pair $(r, m)$
consisting of a run $r$ and a time $m$. 
We make the standard assumption that time ranges over the natural numbers.
At a point $(r, m)$, the system is in some global state $r(m)$. If $r(m)
= (s_e, s_1,
\ldots , s_n)$, then we take $r_i(m)$ to be $s_i$, agent $i$'s local
state at the point $(r, m)$.  
Note that an agent's local state at  point $(r,m)$ does {\em not}
necessarily encode all the agent's previous local states. 
In some systems, agents have perfect recall,
in the sense that their local state $r_i(m)$ encodes their states
at times $0, \ldots, m-1$, but this need not be generally true.
(See \cite[Chapter 4]{FHMV} for a formal definition and discussion of
perfect recall.)
Formally, a {\em system\/} consists of a set of runs (or executions).   
Let $\P(\R)$ denote the points in a system $\R$.

The runs and systems framework is compatible with 
many other standard approaches for representing and reasoning about systems.
For example, the runs might be event traces generated by a CSP process 
(see Section~\ref{sec:CSP}), they
might be message-passing sequences generated by a security protocol,
or they might
be generated from the strands in a strand space \cite{HalPuc01,FHG99}. 
The approach is rich enough to accommodate a variety of system representations.

Another important advantage of the framework is that it it is
straightforward to define formally what
an agent knows at a point in a system.
Given a system $\R$, let $\K_i(r,m)$ be the set of points in $\P(\R)$ 
that $i$ thinks are possible at $(r,m)$, i.e.,
\[ \K_i(r,m) = \{ (r',m') \in \P(\R) : r'_i(m') = r_i(m) \}. \]
Agent $i$ knows a fact
$\phi$ at a point $(r,m)$ if $\phi$ is true at all points in $\K_i(r,m)$.
To make this intuition precise, we need to be able to assign truth
values to basic formulas in a system. We assume that
we have a set $\Phi$ of primitive propositions, which we can think of as
describing basic facts about the system.  
In the context of security protocols,
these might be such facts as ``the key is $n$'' or ``agent $A$ sent
the message $m$ to $B$''. An interpreted system $\I$ consists of a pair
$(\R,\pi)$, where $\R$ is a system and $\pi$ is an 
{\em interpretation}, which assigns to each primitive
proposition
in $\Phi$ a truth value at each point.
Thus, for every $p\in\Phi$ and 
point $(r,m)$ in $\R$, we have $(\pi(r,m))(p)\in \{\bf
true, \bf false\}$. 

We can now define what it means for a formula $\phi$ to be true at a point
$(r, m)$ in an interpreted system $\I$, written $(\I, r, m)\sat\phi$, by
induction on the structure of formulas:
\begin{itemize}
\item $(\I,r,m) \sat p$ iff $(\pi(r,m))(p) = {\bf true}$
\item $(\I,r,m) \sat \neg\phi$ iff $(\I,r,m) \not\sat \phi$
\item $(\I,r,m) \sat \phi \wedge \psi$ iff $(\I,r,m) \sat \phi$ and $(\I,r,m) \sat \psi$ 
\item $(\I, r, m) \sat K_i\phi ~\mbox{iff}~ (\I , r', m') \sat\phi$ for
all $(r',m') \in \K_i(r,m)$
\end{itemize}
As usual, we write $\I \sat \phi$ if $(\I,r,m) \sat \phi$ for all points
$(r,m)$ in $\I$.

\section{Defining Anonymity Using Knowledge}\label{sec:anonymity}

\subsection{Information-Hiding Definitions}\label{sec:infohide}

Anonymity is one example of an
information-hiding requirement. Other information-hiding requirements
include noninterference, privacy, confidentiality, secure message-sending,
and so on. These requirements are similar, and sometimes they overlap.
Noninterference, for example, requires a great deal to be hidden, and
typically implies privacy, anonymity, etc., for the classified user
whose state is protected by the noninterference requirement.

In an earlier paper \cite{HalOn02}, we looked at 
requirements of {\em total secrecy}
in multiagent systems.
Total secrecy
basically requires that in a 
system with ``classified'' and ``unclassified'' users, the unclassified 
users should never be able to infer the actions or the local states of
the unclassified users. For secrecy, the ``what needs to be hidden'' 
component of information-hiding is extremely restrictive: 
total
secrecy requires
that absolutely everything that a classified user does must be hidden.
The ``how well does it need to be hidden'' component depends on the
situation. Our definition of secrecy says
that for any {\em nontrivial\/}
fact $\phi$ (that is, one that is not already valid)
that depends only the state of the classified or high-level
agent, the formula $\neg K_j \phi$ must be valid.
(See 
our earlier paper
for more discussion of this definition.)
Semantically, this means that whatever the high-level user does, there
exists some run where the low-level user's view  of the system is the 
same, but the high-level user did something different.
Our nonprobabilistic definitions are fairly
strong (simply because secrecy requires that so much be hidden).
The probabilistic definitions 
we gave
require even more: not only 
can the agent not learn any new classified
fact, but he also cannot learn anything about the probability of any such
fact. (In other words, if an agent initially assigns a classified 
fact $\phi$ a probability $\alpha$
of being true, he always assigns $\phi$ that probability.)
It would be perfectly natural,
and possibly quite interesting, to
consider definitions of secrecy that do not require so much to be hidden  
(e.g., by allowing some classified information to be 
declassified~\cite{zdancewic01}),
or to discuss definitions that do not require such strong secrecy (e.g., by
giving definitions that were stronger than the nonprobabilistic definitions
we gave, but not quite so strong as the probabilistic definitions).

\subsection{Defining Anonymity}\label{sec:dfns}

The basic intuition behind anonymity is that {\em actions} should be 
divorced from the {\em agents} who perform them, for some set of {\em observers}.
With respect to the basic information-hiding framework outlined above, the
information that needs to be hidden is the identity of the agent (or set of
agents) who perform a particular action. Who the information needs to be
hidden from, i.e., which observers, depends on the situation. The third
component of information-hiding requirements---how well information needs to
be hidden---will often be the most interesting component
of the definitions of anonymity that we present here.

Throughout the paper, we use the formula $\theta(i,a)$ to represent
``agent $i$ has 
performed action $a$, or will perform $a$ in the future''.%
\footnote{If we want to consider systems that may crash
we may want to consider $\theta'(i,a)$ instead, where
$\theta'(i,a)$ represents ``agent $i$ has performed action $a$, or will
perform $a$ in the future if the system does not crash''.  Since issues
of failure are orthogonal to the anonymity issues that we focus on here,
we consider only the simpler definition in this paper.}
For future reference, let $\delta(i,a)$ 
represent ``agent $i$ has performed action $a$''. 
Note that $\theta(i,a)$ is a fact about the run: if it is true at some
point in a run, it is true at all points in a run (since it is true even
if $i$ performs $a$ at some point in the future).  On the other hand,
$\delta(i,a)$ may be false at the start of a run, and then become true
at the point where $i$ performs $a$.

It is not our goal in this 
paper
to provide a ``correct'' definition of 
anonymity. We also want to avoid giving an encyclopedia of definitions. 
Rather, we give some basic definitions of anonymity 
to show how our framework can be used. We base our choice of definitions
in part on definitions presented in earlier papers,
to make clear how our work relates to previous work, and in part on which
definitions of anonymity we expect to be useful in practice.
We first give
an extremely weak definition, but one that
nonetheless illustrates the basic intuition behind any definition of
anonymity.

\dfn
\label{dfn:minimalanon1}
Action $a$, performed by agent $i$, is
{\em minimally anonymous} with respect to agent $j$ in the interpreted
system $\I$, if
$\I \models \neg K_j [\theta(i,a)]$.
\edfn

This definition makes it
clear what is being hidden ($\theta(i,a)$---the fact that
$i$ performs $a$) and from whom
($j$).  It also describes how well the information is hidden:~it
requires that $j$ not be sure that $i$ actually 
performed, or will perform, the action.
Note that this is a weak information-hiding requirement.
It might be the case, for example, that agent $j$ is certain 
that the action was performed either by $i$, or by at most one or two
other agents, thereby making $i$ a ``prime suspect''. It might also
be the case that $j$ is able to place a very high probability on 
$i$ performing
the action, even though he isn't absolutely certain of
it. (Agent $j$ might know that there is some slight probability that
some other agent $i'$ performed the action, for example.)
Nonetheless,
it should be the case that for any other definition of anonymity we
give, if we want 
to ensure that $i$'s performing action $a$ is to be kept anonymous
as far as observer $j$ is concerned, then 
$i$'s action should be at least minimally anonymous with respect to $j$.

Our definition of $a$ being minimally anonymous with respect to $j$ is
equivalent to the apparently weaker requirement $\I \models
\theta(i,a) \rimp \neg K_j 
[\theta(i,a)]$, which says that if action $a$ is performed by $i$, then
$j$ does not not know it.  
Clearly if $j$ never knows that $a$ is performed by $i$, then $j$ will
never know that $a$ is performed by $i$ if $i$ actually does perform
$a$.  To see that the converse holds, it suffices to note that if $i$
does not perform $a$, then surely $\neg K_j [\theta(i,a)]$ holds.  
Thus, this definition, like several that will follow,
can be viewed as having the form ``if $i$
performed $a$, then $j$ does not know some appropriate 
fact''.

The definition 
of minimal anonymity
also makes it clear how anonymity
relates to secrecy, 
as defined in our earlier work \cite{HalOn02}.
To explain how, we first need to describe how we defined secrecy in
terms of knowledge.
Given a system $\I$, say that $\phi$ is {\em nontrivial in $\I$\/} if 
$\I \not\sat \phi$, and that 
{\em $\phi$ depends only on the local state of agent
$i$ in $\I$\/} if $\I \sat \phi \rimp K_i \phi$. 
Intuitively, $\phi$ 
is nontrivial in $\I$ if $\phi$ could be
false in $\I$, and $\phi$ depends only on $i$'s local state if $i$
always knows whether or not $\phi$ is true.  
(It is easy to see that 
$\phi$ depends only on the local state of $i$ if $(\I,r,m) \sat \phi$
and $r_i(m) = r'_i(m')$ implies that $(\I,r',m') \sat \phi$.)
According to the definition in \cite{HalOn02}, 
agent $i$ {\em maintains total secrecy with respect to another agent $j$
in system $\I$\/} if
for every 
nontrivial fact $\phi$ that depends 
only
on the local state of $i$, the formula
$\neg K_j \phi$ is valid for the system. 
That is, $i$ maintains total secrecy with respect to $j$ if 
$j$ does not learn anything new about agent $i$'s state.
In general, $\theta(i,a)$ does not depend only on $i$'s local state,
because whether $i$ performs $a$ may depend on whether or not $i$ gets a
certain message from 
some other agent $i'$.
On the other hand, if whether or not $i$
performs $a$ depends only on $i$'s protocol, and the protocol is encoded
in $i$'s local state, then $\theta(i,a)$ depends only on $i$'s local
state.  
If $\theta(i,a)$ does depend only on $i$'s local state and 
$j$ did not know all along that $i$ was going to
perform action $a$ (i.e., if we assume 
that $\theta(i,a)$ is nontrivial), then
Definition~\ref{dfn:minimalanon1} is clearly a special case of the
definition of secrecy.  
In any case, it is in much the same spirit as the definition of
secrecy. 
Essentially, anonymity says that
the fact that agent $i$ has or will perform action $a$ must be hidden
from $j$, 
while 
total secrecy says that all facts that depend on agent $i$ must
be 
hidden from $j$. 

Note that this definition of minimal anonymity is different from the one
we gave in the conference version of this paper \cite{HalOn03}.
There, the definition given used $\delta(i,a)$ rather than
$\theta(i,a)$.  We say that $a$ performed by agent $i$ is minimally
$\delta$-anonymous if Definition~\ref{dfn:minimalanon1} holds, with
$\theta(i,a)$ replaced by $\delta(i,a)$.  
It is easy to see that minimal anonymity implies minimal
$\delta$-anonymity (since $\delta(i,a)$ implies $\theta(i,a)$), but the
converse is not true in general.   For example, suppose that $j$
gets a signal if $i$ is going to perform action $a$ (before $i$ actually
performs the action), but then never finds out exactly when $i$ performs
$a$.  Then minimal anonymity does not hold.  In runs where $i$ performs
$a$, agent $j$ knows that $i$ will perform $a$ when he gets the signal.
On the other hand, minimal $\delta$-anonymity does hold, because $j$
never knows when $i$ performs $a$.  
In this situation, minimal anonymity seems to capture our intuitions of
what anonymity should mean better than minimal $\delta$-anonymity does.

The next definition of anonymity we give is much stronger. 
It requires that if some agent $i$ performs
an action anonymously with respect to another agent $j$, then $j$ must
think it possible that the action could have been performed by {\em any}
of the agents (except for $j$). 
Let $P_j \phi$ be an abbreviation for 
$\neg K_j \neg \phi$. The operator $P_j$ is the dual of $K_j$; 
intuitively, $P_j \phi$ means ``agent $j$ thinks that $\phi$ is possible''.

\dfn
\label{dfn:totalanon1}
Action $a$, performed by agent $i$, is {\em totally anonymous} 
with respect to $j$ in the interpreted system $\I$ if 
\[ 
\I \models \theta(i,a) \Rightarrow \bigwedge_{i' \ne j} P_j
 [\theta(i',a)].
\]
\edfn

Definition~\ref{dfn:totalanon1} captures the notion that an action is
anonymous if, 
as far as the observer in question is concerned, it could have been
performed by  
anybody in the system. 

Again, in the conference version of the paper, we defined total
anonymity using $\delta(i,a)$ rather than $\theta(i,a)$.  (The same
remark holds for all the other definitions of anonymity that we give,
although we do not always say 
so
explicitly.)  
Let 
total $\delta$-anonymity
be the 
anonymity requirement
obtained when $\theta(i,a)$ is replaced by
$\delta(i,a)$.  It is not hard to show that if agents have perfect
recall (which intuitively means that their local state keeps track of
all the actions they have performed---see \cite{FHMV} for the formal
definition), then total $\delta$-anonymity implies total anonymity. 
This is not true, in general, without perfect
recall, because it might be possible for some agent to know that $i$
will perform action $a$---and therefore that no other agent 
will---but forget this fact by the time that $i$ actually performs
$a$. 
Similarly, total anonymity does not imply total $\delta$-anonymity.
To see why,
suppose that the agents are numbered $1, \ldots, n$, and 
that
an outside observer knows 
that if
$j$ performs 
action $a$, 
then
$j$ will perform it at time $j$.  Then total anonymity may
hold even though total $\delta$-anonymity does not.
For example, at time 3,
although the 
observer 
may consider
it possible that agent 4 will perform the
action (at time 4), he 
cannot
consider it possible that 4 has
already
performed the action,
as required by total $\delta$-anonymity.

Chaum \citeyear{Chaum88} showed that total anonymity could be 
obtained using DC-nets.
Recall that in a DC-net, a group of $n$ users use Chaum's dining
cryptographer's 
protocol (described in the same paper) to achieve anonymous communication.
If we model a  
DC-net as an interpreted multiagent system $\I$ whose agents consist
exclusively of agents 
participating in a single DC-net, then if an agent $i$ sends a message
using the DC-net protocol, that action is totally anonymous. 
(Chaum proves this,
under the assumption that any message could be generated by any user
in the system.)
Note that in the dining cryptographer's example, total anonymity and
$\delta$-total anonymity agree, because who paid is decided 
before the protocol starts. 

It is easy to show that if an
action is totally anonymous, then it must be minimally anonymous as well,
as long as two simple requirements are satisfied. First, there must be
at least 3 agents in the system. 
(A college student with only one roommate can't leave out her dirty
dishes anonymously,  
but a student with at least two roommates might be able to.)
Second, it must be the case that $a$ 
can be performed only once in a given
run of the system. Otherwise, 
it might be possible for $j$ to think that
any agent $i' \not= i$ could have performed $a$, 
but for $j$ to {\em know}
that agent $i$ did, indeed, perform $a$.
For example, consider a system with three agents besides $j$.  Agent $j$
might know 
that all three of the other agents performed action $a$.  In that case, in
particular, $j$ knows that $i$ performed $a$, so action $a$ performed
by $i$ is not minimally anonymous with respect to $j$, but is totally anonymous.
We anticipate that this assumption will typically be met in practice.  
It is certainly consistent 
with examples of anonymity given in the 
literature. (See, for example,~\cite{Chaum88,schneider96}).
In any case, if it is not met, it is possible to tag 
occurrences of an action (so that we can talk about the $k$th time $a$
is performed). Thus,
we can talk about the $i$th occurrence of an action being
anonymous. Because the $i$th occurrence of an action can only happen
once in any given run, our requirement is satisfied.

\pro\label{pro:minimally}
Suppose that there are at least three agents in the interpreted system $\I$
and that 
$$\I \models \bigwedge_{i \not= j} \neg [ \theta(i,a) \wedge \theta(j,a) ].$$
If action $a$, performed by agent $i$, is totally anonymous with respect to $j$,
then it is minimally anonymous as well.
\epro

\prf
Suppose that action $a$ is totally anonymous. Because there are three agents in
the system, there is some agent $i'$ other than $i$ and $j$, and by total 
anonymity, $\I \models \theta(i,a) \Rightarrow P_j[\theta(i',a)] $.
If $(\I,r,m) \models \neg \theta(i,a)$, clearly $(\I,r,m) \models \neg K_j[\theta(i,a)]$.
Otherwise, $(\I,r,m) \models P_j[\theta(i',a)]$ by total anonymity. Thus, there
exists a point $(r',m')$ such that $r'_j(m') = r_j(m)$ and $(\I,r',m') \models \theta(i',a)$.
By our assumption, $(\I,r',m') \models \neg \theta(i,a)$, because $i \not= i'$.
Therefore, $(\I,r,m) \models \neg \K_j[\theta(i,a)]$. It follows that $a$
is minimally anonymous with respect to $j$.
\eprf

Definitions~\ref{dfn:minimalanon1} and~\ref{dfn:totalanon1} are
conceptually similar, even 
though the latter definition is much stronger. 
Once again, there is a set of formulas that an
observer is not allowed to know. With the earlier definition, there is only one
formula in this set: $\theta(i,a)$. As long as $j$ doesn't know that $i$ performed
action $a$, this requirement is satisfied. With total anonymity, there
are more formulas that $j$
is not allowed to know: they
take the form $\neg \theta(i',a)$. Before,
we could guarantee only
that $j$ did not know that $i$ did the action; here, for many agents $i'$,
we guarantee that $j$ does not know that $i'$ did {\em not} do the action. The
definition is made slightly more complicated by the implication, which restricts
the conditions under which $j$ is not allowed to know $\neg
\theta(i',a)$. (If $i$ 
didn't actually perform the action, we don't care what $j$ thinks,
since we are concerned only with anonymity with respect to $i$.) 
But the basic idea is the same.

Note that total anonymity does {\em not\/} necessarily follow from total
secrecy, because  
the formula $\neg \theta(i',a)$, for $i' \ne i$, does not, in general,
depend only on the local 
state of $i$. It is therefore perfectly consistent
with the definition of total secrecy 
for $j$ to learn this fact,
in violation of total anonymity.
(Secrecy, of course, does not follow from anonymity, because secrecy
requires that 
many more facts be hidden than simply whether $i$ performed a given action.)

Total anonymity is a very strong requirement.
Often, an action will not be totally anonymous, but
only anonymous up to some set of agents who could have performed the action.
This situation merits a weaker definition of anonymity.
To be more precise, let $I$ be the set of all agents of the system and 
suppose that we have some 
set $I_A \subseteq I$---an ``anonymity set'', using the terminology
of Chaum~\citeyear{Chaum88} and 
Pfitzmann and K\"ohntopp~\citeyear{terminology}---%
of agents who can perform some action. We can define anonymity in
terms of this set.

\dfn
\label{dfn:anonset1}
Action $a$, performed by agent $i$, is 
{\em anonymous up to $I_A \subseteq I$} with respect to $j$ if 
\[ \I \models \theta(i,a) \Rightarrow \bigwedge_{i' \in I_A} P_j [\theta(i',a)]. \]
\edfn
In the anonymous message-passing system Herbivore~\cite{goel02}, users are
organized into {\em cliques} $C_1, \ldots, C_n$, each of which uses the dining cryptographers
protocol \cite{Chaum88} for anonymous message-transmission. If a user wants to send
an anonymous message, she can do so through her clique. 
Herbivore claims
that any user $i$ is
able to send a message anonymously up to $C_j$, where $i \in C_j$. As the size of
a user's clique varies, so does the strength of the anonymity guarantees provided
by the system.

In some situations, 
it is not necessary that there be a fixed 
anonymity set,
as in Definition~\ref{dfn:anonset1}.  It suffices that, at all times,
there {\em exists} some 
anonymity set
with at least, say, $k$ agents. This leads to a definition of $k$-anonymity.

\dfn
\label{dfn:anonset2}
Action $a$, performed by agent $i$, is 
{\em $k$-anonymous} with respect to $j$ if
\[ \I \models \theta(i,a) \Rightarrow \bigvee_{\{I_A: |I_A| = k\}}
\bigwedge_{i' \in I_A} P_j 
[\theta(i',a)]. \] 
\edfn

This definition says that at any point $j$ must think it possible that
any of at least $k$ agents might perform, or have performed, the action.
Note that the set of $k$ agents might be different in different runs,
making this condition strictly weaker than anonymity up to a particular
set of size $k$.

A number of systems have been proposed that provide $k$-anonymity for
some $k$.
In the anonymous communications network protocol recently proposed by
von Ahn, Bortz, and Hopper \cite{hopper03}, users can 
send 
messages with guarantees of $k$-anonymity. In the system $P^5$
(for 
``Peer-to-Peer Personal Privacy Protocol'') \cite{sherwood02}, 
users join a logical broadcast
tree that provides anonymous communication, and users can choose what
level 
of $k$-anonymity they want, given that $k$-anonymity for a higher value of $k$
makes communication more inefficient.
Herbivore \cite{goel02} provides anonymity using 
cliques of DC-nets.  If
the system guarantees that
the cliques all have a size of at least $k$, 
so that regardless of clique
composition, there are at least $k$ users capable of sending any
anonymous message, then Herbivore guarantees $k$-anonymity.

\subsection{A More Detailed Example: Dining Cryptographers}\label{sec:diningexample}

A well-known example of anonymity in the computer security literature
is Chaum's ``dining cryptographers problem''~\cite{Chaum88}. In the 
original description of this problem, three cryptographers sit down to
dinner and are informed by the host that someone has already paid the bill 
anonymously. The cryptographers decide that the bill was paid either by
one of the three people in their group, or by an outside agency such as
the NSA. They want to find out which of these two situations is the actual
one while preserving the anonymity of the cryptographer who (might have) paid.
Chaum provides a protocol that the cryptographers can use to solve this 
problem. To guarantee that it works, however, it would be nice to check
that anonymity conditions hold. Assuming we have a system that includes
a set of three cryptographer agents $C = \{0, 1, 2\}$, 
as well as an outside 
observer agent $o$, the protocol should guarantee that for each agent $i
\in C$, 
and each agent $j \in C - \{i\}$,
the act of paying is anonymous up to $C - \{j\}$ with respect to $j$.
For an outside observer $o$, 
i.e., an agent other than one of three cryptographers,
the protocol should guarantee that for
each agent $i \in C$, the protocol is anonymous up to $C$ with respect to
$o$. This can be made precise using our definition of anonymity up to a set.

Because the requirements are symmetric for each of the three cryptographers,
we can 
describe the anonymity specification compactly by naming
the agents using 
modular arithmetic. We use $\oplus$ to denote addition mod 3. 
Let the interpreted system $(\I = (\R,\pi)$ represent the possible runs
of one instance of the 
dining cryptographers protocol, where the interpretation
$\pi$ interprets formulas of the form 
$\theta(i,\mbox{``paid''})$
in the obvious way. 
The following knowledge-based requirements comprise the anonymity portion of the protocol's
specification, for each agent $i \in C$:
$$
\begin{array}{ll}
\I \models \theta(i,\mbox{``paid''}) &\Rightarrow 
  P_{i \oplus 1} \theta(i \oplus 2,\mbox{``paid''}) 
   \wedge~ P_{i \oplus 2} \theta(i \oplus 1,\mbox{``paid''})\\ 
  &\wedge~ P_o \theta(i \oplus 1,\mbox{``paid''}) 
   \wedge~ P_o \theta(i \oplus 2,\mbox{``paid''}). 
\end{array}
$$
This means that if a cryptographer paid, then each of the other
cryptographers must think it possible that the third cryptographer
could have paid. In addition, an outside observer must think it 
possible that either of the other two cryptographers could have paid.

\section{Probabilistic Variants of Anonymity}\label{sec:probanon}

\subsection{Probabilistic Anonymity}

All of the definitions presented in Section~\ref{sec:anonymity} were nonprobabilistic.
As we mentioned in the introduction,
this is a serious problem for the ``how well is information hidden'' component of
the definitions. For all the definitions we gave, it was necessary only that 
observers think it {\em possible} that multiple agents could have performed the
anonymous action. However, an event that is possible may nonetheless be 
extremely unlikely. Consider our definition of total anonymity 
(Definition~\ref{dfn:totalanon1}). It states that an action performed by $i$
is totally anonymous if the observer $j$ thinks it could have been performed
by any agent other than $j$. This may seem like a strong requirement, but if there
are, say, $102$ agents, 
and $j$ can determine that $i$ performed action $a$
with probability $0.99$ and that each of the other agents performed action
$a$ with probability $0.0001$, agent $i$ might not be very happy with the
guarantees provided by total anonymity.
Of course, the appropriate notion of anonymity will depend on the
application:
$i$ might be content to know that no agent can {\em prove} that she
performed the anonymous action. In that case, it
might suffice for the action to be only
minimally anonymous. 
However, in many other cases, an agent might want a more quantitative,
probabilistic guarantee that it will be considered reasonably
likely that other agents could have performed the action. 

Adding probability to the runs and systems framework is straightforward.
The approach we use goes back to \cite{HT}, and was also used in our
work on secrecy \cite{HalOn02}, so we just briefly review the relevant
details here.
Given a system $\R$, suppose we have a probability measure $\mu$ on the runs
of $\R$. The pair $(\R, \mu)$ is a {\em probabilistic system}. For simplicity,
we assume that every subset of $\R$ is measurable. We are interested
in the probability that an agent assigns to an event at the point $(r,m)$.
For example, we may want to know that at the point $(r,m)$, 
observer $i$ places a probability of $0.6$
on $j$'s having performed some particular action. We want to condition
the probability $\mu$ on $\K_i(r,m)$, the information that $i$ has
at the point $(r,m)$. The problem is that $\K_i(r,m)$ is a set of {\em
points}, while $\mu$ is a probability on {\em runs}.   
This problem is dealt with as follows.

Given a set $U$ of points, let $\R(U)$ consist of the runs in $\R$ going through
a point in $U$. That is, 
$$\R(U) = \{r \in \R: (r,m) \in U \mbox{ for some $m$}\}.$$
The idea will be to condition $\mu$ on $\R(\K_i(r,m))$ rather than on 
$\K_i(r,m)$.  To make sure that conditioning is well defined, 
we assume that $\mu(\R(\K_i(r,m))) > 0$ for each agent $i$, run $r$, 
and time $m$. That is, $\mu$ assigns positive probability to the set of
runs in $\R$ compatible with what happens in run $r$ up to time $m$,
as far as agent $i$ is concerned.  

With this assumption, we can define a measure $\mu_{r,m,i}$ on the
points in $\K_i(r,m)$ as follows.  If $\S \subseteq \R$, define
$\K_i(r,m)(\S)$ to be the set of points in $\K_i(r,m)$ that lie on runs
in $\S$; that is,
$$\K_i(r,m)(\S) = \{(r',m') \in \K_i(r,m): r' \in \S\}.$$
Let $\F_{r,m,i}$, the measurable subsets of $\K_i(r,m)$ (that is, the
sets to which $\mu_{r,m,i}$ assigns a probability), consist of all sets of the
form $\K_i(r,m)(\S)$, where $\S \subseteq \R$.  Then define 
$\mu_{r,m,i}(\K_i(r,m)(\S)) = \mu(\S \mid \R(\K_i(r,m))$.
It is easy to check that $\mu_{r,m,i}$ is a probability measure,
essentially defined by conditioning.

Define a {\em probabilistic interpreted system\/} $\I$ to be a tuple
$(\R,\mu,\pi)$, where $(\R,\mu)$ is a probabilistic system.
In a probabilistic interpreted system, we can give semantics to
syntactic statements of probability. 
Following \cite{FHM}, we will be most interested in formulas of the form
$\Pr_i (\phi ) \leq \alpha$ 
(or similar formulas with $>$, $<$, or $=$
instead of $\leq$). Intuitively, a 
formula such as $\Pr_i (\phi) \leq \alpha$ is true 
at a point $(r,m)$ if, according to $\mu_{r,m,i}$, the probability that
$\phi$ is true is at 
most
$\alpha$.  More formally,
$  (\I,r,m) \models {\Pr}_i(\phi) \leq \alpha $ if
\[  \mu_{r,m,i}(\{(r',m') \in \K_i(r,m) : (\I,r',m') \models \phi \})
\leq \alpha. \] 
Similarly, we can give semantics to $\Pr_i(\phi) < \alpha$ and
$\Pr(\phi) = \alpha$, as well as conditional formulas such as
$\Pr(\phi \, | \, \psi) \le \alpha$.
Note that although these formulas talk
about probability, they are either true or false at a given state.

It is straightforward to define probabilistic notions of anonymity in
probabilistic systems.  
We can think of Definition~\ref{dfn:minimalanon1}, for example, as saying
that $j$'s probability that
$i$ 
performs
the anonymous action $a$ must be less than 1
(assuming that every nonempty set has positive probability).
This can be generalized
by specifying some $\alpha \leq 1$
and requiring that the probability of $\theta(i,a)$ be less than $\alpha$.

\dfn
\label{dfn:alphaanon1}
Action $a$, performed by agent $i$, is {\em $\alpha$-anonymous}
with respect to agent $j$ if 
$\I \models \theta(i,a) \Rightarrow {\Pr}_j [\theta(i,a)] < \alpha$. 
\edfn

Note that if we replace $\theta(i,a)$ by $\delta(i,a)$ in
Definition~\ref{dfn:alphaanon1}, the resulting notion might not be well
defined.  The problem is 
that the set
$$\{(r',m') \in \K_i(r,m): (\I,r',m') \sat \delta(i,a) \}$$ 
may not be
measurable; it may not have the form $\K_i(r,m)(\S)$ for some $\S
\subseteq \R$.  The problem does not arise if $\I$ is a {\em
synchronous\/} sytem (in which case $i$ knows that time, and all the
points in $\K_i(r,m)$ are of the form $(r',m)$), but it does arise if
$\I$ is asynchronous.  We avoid this technical problem by working with
$\theta(i,a)$ rather than $\delta(i,a)$.

Definition~\ref{dfn:alphaanon1}, unlike
Definition~\ref{dfn:minimalanon1},
includes an implication involving $\theta(i,a)$. It is easy
to check that Definition~\ref{dfn:minimalanon1} does not change
when such an implication is added; intuitively, if $\theta(i,a)$
is false then $\neg K_j[\theta(i,a)]$ is trivially true. 
Definition~\ref{dfn:alphaanon1}, however, would change if we removed
the implication,
because it might be possible for $j$ to have a high probability
of $\theta(i,a)$ even though it isn't true. We include the implication
because without it, we place constraints on what $j$ thinks about 
$\theta(i,a)$ even if $i$ has not performed the action $a$ and
will not
perform it in the future. Such a requirement,
while interesting,
seems more akin to ``unsuspectibility'' than to anonymity.

Two of the notions of probabilistic anonymity considered by 
Reiter and Rubin \citeyear{reiter98} in the context of their Crowds
system can be understood in terms of $\alpha$-anonymity.
Reiter and Rubin 
say that a sender has {\em probable innocence}
if, from an observer's point of view, the sender ``appears no more likely to be the
originator than to not be the originator''. This is simply 0.5-anonymity. 
(Under reasonable assumptions, Crowds provides 0.5-anonymity for Web requests.)
Similarly, a sender has {\em possible innocence} if, from an observer's point of
view, ``there is a nontrivial probability that the real sender is someone else''.
This corresponds to minimal anonymity (as defined in Section~\ref{sec:dfns}), or
to $\epsilon$-anonymity for some nontrivial value of $\epsilon$.

It might seem at first that Definition~\ref{dfn:alphaanon1} should be the only
definition of anonymity we need: as 
long as $j$'s probability of 
$i$ performing the action is low
enough, $i$ should have nothing to worry about. 
However, with further thought, it is not hard to see that this is not
the case.  

Consider a scenario where there are 1002 agents, and where $\alpha = 0.11$.
Suppose that the probability, according to Alice, that Bob 
performs
the
action is $.1$, but that her probability that any of the other $1000$ agents 
performs 
the action is $0.0009$ (for each agent). Alice's probability that Bob 
performs
the
action is small, but her probability that anyone else 
performs
it is
more than three orders of magnitude smaller. 
Bob is obviously the prime suspect.

This concern was addressed by Serjantov and Danezis~\citeyear{Serj02}
in their paper
on information-theoretic definitions of anonymity. They consider
the probability that each agent in an anonymity set is the sender
of some anonymous message, and use entropy to quantify the amount of
information that the system is leaking;
Diaz et al.~\citeyear{Diaz02} and 
Danezis~\citeyear{danezis:pet2003} use similar techniques.
In this paper we are not
concerned with quantitative measurements of anonymity, but we
do agree that it is worthwhile to consider stronger
notions of anonymity than the nonprobabilistic definitions, or
even $\alpha$-anonymity, can provide.
We hope to examine quantitative definitions in future work.

The next definition strengthens
Definition~\ref{dfn:alphaanon1} in the way that
Definition~\ref{dfn:totalanon1} strengthens Definition~\ref{dfn:minimalanon1}.
It requires that no agent in the 
anonymity
set be a more
likely suspect than any other.

\dfn
\label{dfn:probanonset1}
Action $a$, performed by agent $i$, is {\em strongly 
probabilistically anonymous up to 
$I_A$\/}  with respect to 
agent $j$ if for each $i' \in I_A$,
\[ \I \models \theta(i,a) \Rightarrow {\Pr}_j [\theta(i,a)] = {\Pr}_j [\theta(i',a)]. \]
\edfn

Depending on the size of $I_A$, this definition can be extremely
strong. It does not 
state simply that for all agents in $I_A$, the observer must think it is
reasonably likely 
that the agent could have performed the action; it also says that the
observer's probabilities must be the same for each such agent. 
Of course, we could weaken the definition somewhat by not requiring that
all the probabilities be equal, but by instead requiring
that they be approximately equal
(i.e., that their difference be small or that their ratio be close to 1).  
Reiter and Rubin \citeyear{reiter98}, for example, say that the sender of a
message is {\em beyond suspicion} if she ``appears no more likely to
be the originator of that message than any other potential sender in the system''.
In our terminology, $i$ is beyond suspicion with respect to $j$ if for
each $i' \in I_A$,
\[ \I \models \theta(i,a) \Rightarrow {\Pr}_j [\theta(i,a)] \leq {\Pr}_j [\theta(i',a)]. \]
This is clearly weaker than strong probabilistic anonymity, but still a very
strong requirement,
and perhaps more reasonable, too.
Our main point is that a wide variety of properties can be expressed
clearly and succinctly in our framework.

\subsection{Conditional Anonymity}

While we have shown that many useful notions of anonymity---including many definitions
that have already been proposed---can be expressed
in our framework, we claim
that there are some important intuitions that have not yet been captured.
Suppose, for example, that someone
makes a \$5,000,000 donation to Cornell University. It is
clearly not the case 
that
everyone is equally likely, or even almost equally likely, to
have made the donation.
Of course, we could take the 
anonymity
set $I_A$ to
consist of those people who might be in a position to make such a large donation,
and insist that they all be considered equally likely. Unfortunately,
even that is unreasonable: a priori, some of them may already have known
connections to Cornell, and thus 
be considered far more likely to
have made the donation. All that an anonymous donor
can reasonably expect is that nothing an
observer learns from his interactions with the environment (e.g., 
reading the newspapers, noting when the donation was made, etc.)~will
give him more information about the identity of the donor than he
already had.

For another example, consider a conference or research journal 
that provides anonymous reviews to researchers who submit their papers for 
publication. 
It is unlikely that the review process provides anything like
$\alpha$-anonymity for a small $\alpha$, or strongly probabilistic
anonymity up to some reasonable set.
When 
the preliminary version of
this paper, for example, was accepted by 
the Computer Security Foundations Workshop,
the acceptance notice included three reviews that were, in our terminology,
anonymous up
to the program committee. That is, any one of the reviews we received
could have been written by any of the members of the program committee. However,
by reading some of the reviews, we were able to make fairly good guesses 
as to
which committee members had provided which reviews, based on our knowledge
of the specializations of the various members, and based on the
content of the reviews themselves. 
Moreover, we had a fairly good idea of which committee members
would provide reviews of our paper even before we received the
reviews.
Thus, it seems unreasonable to hope that the review process would
provide strong probabilistic anonymity (up to the program committee),
or even some weaker variant of probabilistic anonymity.
Probabilistic anonymity would require the reviews to convert our prior
beliefs, according to which
some program committee members were more likely than others to be
reviewers 
of our paper, to posterior beliefs 
according to which all program committee members were equally likely! 
This does not seem at all reasonable.  
However, the reviewers might hope that 
that the process did not
give us any more information than we already had.

In our paper on secrecy \cite{HalOn02},
we tried to capture 
the intuition that, when an unclassified user interacts with a secure
system, she does not learn anything about any classified user 
that she didn't already know.  We did 
this formally by requiring that, for any three 
points $(r,m)$, $(r',m')$, and $(r'',m'')$, 
\begin{equation}\label{secrecy}
\mu_{(r,m,j)}(\K_i(r'',m'')) = \mu_{(r',m',j)}(\K_i(r'',m'')).
\end{equation}
That is, whatever the unclassified user $j$ sees, her probability of
any particular classified state will remain unchanged.

When defining anonymity, we are not concerned with protecting 
all information about some agent $i$, but rather the fact that $i$
performs
some particular action $a$. 
Given a probabilistic system $\I = (\R,\pi,\mu)$ and a formula $\phi$,
let $e_r(\phi)$ 
consist of the set of runs $r$ such 
that $\phi$ is true at some point in $r$, and let $e_p(\phi)$ be the set
of points where $\phi$ 
is true.  That is
$$\begin{array}{l}
e_r(\phi) = \{ r : \exists m ( (\I,r,m) \models \phi) \}, \\
e_p(\phi) = \{ (r,m) : (\I,r,m) \models \phi \}.
\end{array}
$$
The most obvious analogue to (\ref{secrecy}) is the requirement that,
for all points $(r,m)$ and $(r',m')$, 
$$\mu_{(r,m,j)}(e_p(\theta(i,a))) = \mu_{(r',m',j)}(e_p(\theta(i,a))).$$
This definition says that $j$ never learns anything about the
probability that $i$ 
performed 
performs
$a$: 
she always ascribes the same probability to this event.
In the context of our anonymous donation example, 
this would say that the probability (according to $j$) of $i$ donating
\$5,000,000 to Cornell is the same at all times.

The problem with this definition is that it does not allow 
$j$ to learn that {\em someone\/} donated \$5,000,000 to Cornell.  That is, before
$j$ learned that someone donated \$5,000,000 to Cornell, $j$ may have
thought it was unlikely that anyone would donate that much money to
Cornell.  We cannot expect that $j$'s probability of $i$ donating
\$5,000,000 would be the same both before and after learning that
someone made a donation. 
We want to give a definition of conditional anonymity that
allows observers to learn that an action has been performed,
but that protects---as much as possible, given the system---the
fact that some particular agent 
performed 
performs
the action. If, on
the other hand, the anonymous action has not been performed,
then the observer's probabilities do not matter.

Suppose that $i$ wants to perform action $a$, and wants conditional anonymity
with respect to $j$. Let
$\theta(\overline{\jmath},a)$ represent the fact that $a$ has been performed
by some agent other than 
$j$, i.e., $\theta(\overline{\jmath},a) =
\lor_{i' \ne j} \theta(i',a)$.
The definition of conditional anonymity says that $j$'s prior probability 
of $\theta(i,a)$ given $\theta(\overline{\jmath},a)$ must be the
same as his posterior probability of $\theta(i,a)$ at points where $j$ knows 
$\theta(\overline{\jmath},a)$, i.e., at points where $j$
knows that someone other than $j$ has performed 
(or will perform)
$a$.
Let $\alpha = \mu(e_r(\theta(i,a)) \mid e_r(\theta(\overline{\jmath},a)))$. This is the prior 
probability that $i$ has performed $a$, given that somebody other than
$j$ has. 
Conditional anonymity says that at any point where 
$j$ knows that someone other than $j$ 
performs
$a$, 
$j$'s probability of $\theta(i,a)$ must be $\alpha$. 
In other words, $j$ shouldn't be able to learn anything more
about who 
performs
$a$ (except that 
somebody does) 
than he know before he began interacting with the system in the first place.

\dfn\label{dfn:conditionalanon}
Action $a$, performed by agent $i$, is {\em conditionally anonymous} with
respect to $j$ in the probabilistic system $\I$ if 
$$
\I \models K_j\theta(\overline{\jmath},a) \Rightarrow
{\Pr}_j(\theta(i,a)) = \mu(e_r(\theta(i,a)) \mid 
e_r(\theta(\overline{\jmath},a))). 
$$
\edfn
Note that if only one agent ever performs $a$, then
$a$ is trivially conditionally anonymous with respect to $j$, but may
not be minimally anonymous with respect to $j$.  
Thus, conditional anonymity does not necessarily imply minimal anonymity.

In Definition~\ref{dfn:conditionalanon}, we implicitly
assumed that agent $j$ was allowed to learn that someone other than $j$
performed action $a$; anonymity is
intended to hide {\em which\/} agent performed $a$, given that somebody did.
More generally, we believe that
we need to consider anonymity
with respect to what an observer
is allowed to learn.  
We might want to specify, for example, that an observer is allowed to know
that a donation was made, and for how much, or to learn the contents of a
conference paper review. The following definition lets us do this formally.

\dfn\label{dfn:conditionalanon1}
Action $a$, performed by agent $i$, is {\em conditionally anonymous} with
respect to $j$ and $\phi$ in the probabilistic system $\I$ if 
$$ \I \models K_j \phi \Rightarrow
{\Pr}_j(\theta(i,a)) = \mu(e_r(\theta(i,a)) \mid e_r(\phi)). $$
\edfn
Definition~\ref{dfn:conditionalanon} is clearly the special case of
Definition~\ref{dfn:conditionalanon1} where $\phi =
\theta(\overline{j},a)$.  
Intuitively, both of these definitions say that once an observer learns some
fact $\phi$ connected to the fact $\theta(i,a)$, we require that she
doesn't learn anything else that might change her probabilities of
$\theta(i,a)$.

\subsection{Example: Probabilistic Dining Cryptographers}

Returning the dining cryptographers problem, suppose that it is well-known
that one of the three cryptographers at the table is much more generous than
the other two, and therefore more likely to pay for dinner. 
Suppose, for example, that the probability 
measure on the set of runs where the generous
cryptographer has paid is 0.8, given that one of the cryptographers paid for
dinner, and that it is 0.1 for each of the other two cryptographers.
Conditional anonymity for each of the three cryptographers with respect to
an outside observer means that when such observer learns that one of the
cryptographers has paid for dinner, his probability that any of the three
cryptographers paid should remain 0.8, 0.1, and 0.1. 
If the one of the thrifty cryptographers paid, the generous
cryptographer 
should think that there is a probability of $0.5 = 0.1 / (0.1 + 0.1)$ 
that either of the others paid.
Likewise, if the generous cryptographer paid, each of the others should
think that 
there is a probability of $0.8 / (0.8 + 0.1)$ that the generous
cryptographer paid and a 
probability of $0.1 / (0.8 + 0.1)$ that the other thrifty cryptographer
paid.
We can similarly calculate all the other relevant probabilities.

More generally, suppose that
we have an intepreted probabilistic system 
$(\R,\mu,\pi)$ that represents
instances of the dining cryptographers protocol, where the interpretation $\pi$
once again interprets formulas of the form 
$\theta(i,\mbox{``paid''})$ and $\theta(\overline{\jmath},\mbox{``paid''})$
in the obvious way, and where the formula $\gamma$ is true if one of the cryptographers
paid.  (That is, $\gamma$ is equivalent to $\bigvee_{i \in \{0,1,2\}} \theta(i,\mbox{``paid''})$.)
For any cryptographer $i \in \{ 0, 1, 2 \}$, let $\alpha(i)$ be the prior probability that $i$ paid,
given that somebody else did. That is, let
$$\alpha(i) = \mu(e_r(\theta(i,\mbox{``paid''})) \mid e_r(\gamma)).$$
In the more concrete example given above, if $0$ is the generous cryptographer, we would
have $\alpha(0) = 0.8$ and $\alpha(1) = \alpha(2) = 0.1$.

For the purposes of conditional probability with respect to an agent $j$, 
we are interested in the probability that some agent $i$ paid, given that somebody
other than $j$ paid. Formally, for $i \not= j$, let
$$\alpha(i,j) = \mu(e_r(\theta(i,\mbox{``paid''})) \mid e_r(\theta(\overline{\jmath},\mbox{``paid''}))).$$
If an observer $o$ is not one of the three cryptographers, than $o$ didn't pay, 
and we have $\alpha(i,o) = \alpha(i)$. Otherwise, if $i,j \in \{0,1,2\}$,
 we can use conditioning to compute $\alpha(i,j)$:
\[ \alpha(i,j) = \frac{\alpha(i)}{\alpha(j \oplus 1) + \alpha(j \oplus 2)}. \]
(Once again, we make our definitions and requirements more compact
by using modular arithmetic, where $\oplus$ denotes addition mod 3.) 

The following formula captures the requirement of conditional
anonymity in the dining cryptographer's protocol, for each cryptographer
$i$, with respect to the other cryptographers and any outside observers.

$$
\begin{array}{ll}
\I \models & \left[ K_{i \oplus 1} \theta(\overline{i \oplus 1},\mbox{``paid''}) \Rightarrow {\Pr}_{i \oplus 1}(\theta(i,\mbox{``paid''})) = \alpha(i,i \oplus 1) \right] \wedge \\
	& \left[ K_{i \oplus 2} \theta(\overline{i \oplus 2},\mbox{``paid''}) \Rightarrow {\Pr}_{i \oplus 2}(\theta(i,\mbox{``paid''})) = \alpha(i,i \oplus 2) \right] \wedge \\
	& \left[ K_{o} \theta(\overline{o},\mbox{``paid''}) \Rightarrow
	{\Pr}_{o}(\theta(i,\mbox{``paid''})) = \alpha(i,o) \right]. 
\end{array}
$$

Chaum's original proof that the dining cryptographers protocol provides
anonymity actually proves conditional anonymity in this general setting.
Note that if the probability that one of the cryptographers will pay is 1,
that cryptographer will have conditional anonymity even though he doesn't even
have minimal anonymity.

\subsection{Other Uses for Probability}

In the previous two subsections, we have emphasized how probability can be used
to obtain definitions of anonymity stronger than those presented in 
Section~\ref{sec:anonymity}. However, probabilistic systems can also be used
to define interesting ways of weakening those definitions. Real-world
anonymity systems do not offer absolute guarantees of anonymity such as those
those specified by our definitions. Rather, they guarantee that a user's anonymity
will be protected {\em with high probability}. In a given run, a user's anonymity
might be protected or corrupted. If the probability of the event that a user's
anonymity is corrupted is very small, i.e., the set of runs where her anonymity is
not protected is assigned a very small probability by the measure $\mu$, this
might be enough of a guarantee for the user to interact with the system.

Recall that we said that $i$ maintains total anonymity with respect to $j$
if the fact $\phi = \theta(i,a) \Rightarrow \bigwedge_{i' \ne j} P_j [\theta(i',a)]$
is true at every point in the system. Total anonymity is compromised in a run $r$ 
if at some point $(r,m)$, $\neg \phi$ holds. Therefore, the set of runs where
total anonymity is compromised is simply $e_r(\neg \phi)$, using the notation of
the previous section. If $\mu(e_r(\neg \phi))$ is very small, then $i$ maintains
total anonymity with very high probability. This analysis can obviously be extended
to all the other definitions of anonymity given in previous sections.

Bounds such as these are useful for analyzing real-world systems. The Crowds 
system \cite{reiter98}, for example, uses randomization when routing communication traffic,
so that anonymity is protected with high probability. The probabilistic guarantees provided
by Crowds were analyzed formally by 
Shmatikov \citeyear{shmat02}, using a probabilistic model
checker, and he demonstrates how the anonymity guarantees provided by the Crowds system change
as more users (who may be either honest or corrupt) are added to the system. Shmatikov
uses a temporal probabilistic logic to express probabilistic anonymity properties, so these
properties can be expressed in our system framework.  (It is
straightforward to give semantics to temporal operators in systems; see \cite{FHMV}.)
In any case, Shmatikov's analysis of
a real-world anonymity system is a useful example of how the formal methods that we advocate
can be used to specify and verify properties of real-world systems.

\section{Related Work}\label{sec:related}

\subsection{Knowledge-based Definitions of Anonymity}
As mentioned in the introduction, we are not the first to use knowledge
to handle definitions of security, information hiding, or even anonymity.
Anonymity has been formalized using epistemic logic by Syverson
and 
Stubblebine~\citeyear{syverson99}.  Like us, they use epistemic logic
to characterize a number of 
information-hiding requirements that involve anonymity.
However, the focus of their work is very different from ours.
They describe a logic for reasoning about anonymity
and a number of axioms for the logic.
An agent's knowledge is based, roughly speaking, on
his recent actions and observations, as well as
what follows from his log of system events.
The first five axioms that Syverson and Stubblebine
give are the standard {\bf S5} axioms for knowledge.  There 
are well-known soundness and completeness results relating the
{\bf S5} axiom system to Kripke structure semantics for knowledge
\cite{FHMV}. However, they give many more axioms, and 
they do not attempt to give a semantics for which their axioms are
sound. Our focus, on the other hand, is completely semantic.  We have not tried
to axiomatize anonymity. Rather, we try to give an appropriate semantic
framework in which to consider anonymity.

In some ways, 
Syverson and Stubblebine's model is more detailed than the model used here.
Their logic includes
many formulas that represent various actions and facts, including the
sending and 
receiving of messages, details of encryption and keys, and so on. They also 
make more assumptions about the local state of a given agent,
including details about the sequence of actions that the agent has performed
locally, a log of system events that have been recorded, and a set of facts
of which the agent is aware. While these extra details may accurately
reflect the 
nature of agents in real-world systems, 
they are orthogonal to our concerns here.  In any case,
it would be easy to add
such expressiveness to our model as well, simply by including these details in
the local states of the various agents.

It is straightforward to relate our definitions to those of Syverson and
Stubblebine. They consider facts of the form 
$\phi(i)$, where $i$ is a principal, i.e., an agent. 
They assume that the fact $\phi(i)$ is a single
formula in which a single agent name occurs. Clearly, 
$\theta(i,a)$ is an example of such a formula. 
In fact, Syverson and Stubblebine assume
that if $\phi(i)$ and $\phi(j)$ are both true, then $i = j$. 
For the $\theta(i,a)$ formulas, this means
that $\theta(i,a)$ and $\theta(i',a)$ cannot be simultaneously 
true: at most one agent can perform an action in a given run, 
exactly as in the setup of Proposition~\ref{pro:minimally}.

There is one definition in~\cite{syverson99}
that is especially relevant to our discussion; the other relevant
definitions presented there are similar.
A system is said to satisfy
{\em $(\geq k)$-anonymity} if the following formula is valid for
some observer $o$: 
$$
\begin{array}{c}
\phi(i) \Rightarrow P_o(\phi(i)) \wedge P_o(\phi(i_1)) 
\wedge \cdots \wedge P_o(\phi(i_{k-1})). 
\end{array}
$$
This definition says that if $\phi(i)$ holds, there must be at least $k$ agents, including
$i$, that the observer suspects. (The existential quantification of the agents $i_1, \ldots, i_{n-1}$ is
implicit.)
The definition
is essentially equivalent to our definition of $(k-1)$-anonymity.  
It certainly implies that there are $k-1$ agents other than
$i$ for which $\phi(i')$ might be true. On the other
hand,
if $P_o(\phi(i'))$ is true for $k-1$ agents other than $i$,
then the formula must hold, 
because
$\phi(i) \rimp P_o(\phi(i))$ is valid.

\subsection{CSP and Anonymity}\label{sec:CSP}
A great deal of work on the foundations of computer security has used process
algebras such as CCS and CSP \cite{Mil,hoare85} as the basic system 
framework \cite{focardi01,schneider96b}. Process algebras
offer several advantages: they are simple, they can be used for specifying
systems as well as system properties, and model-checkers 
are available that can be used
to verify properties of systems described using their formalisms.

Schneider and Sidiropoulos~\citeyear{schneider96} use CSP 
both to characterize one type of anonymity and to describe variants of
the dining cryptographers
problem~\cite{Chaum88}. They then use a model-checker to verify 
that their notion of anonymity holds for those variants of the problem.
To describe their approach, we need to outline
some of the basic notation and semantics of CSP. To save space, we give a
simplified treatment of CSP here. 
(See Hoare~\citeyear{hoare85} for a complete description of CSP.) 
The basic unit of
CSP is the {\em event}. Systems are modeled in terms of the events that
they can perform. Events may be built up several components. For example,
``donate.\$5''
might represent a ``donate'' event in the amount of \$5. {\em Processes} are the
systems, or components of systems, that are described using CSP. As a
process
unfolds or executes, various events occur. 
For our purposes, we make the simplifying assumption that a process
is determined by the event sequences it is able to engage in.

We can associate with every process a set of {\em traces}.  
Intuitively, each trace in the set associated with process $P$
represents one sequence of events that might occur during an execution
of $P$.
Informally, CSP event traces correspond to
finite prefixes of runs, except that they do not explicitly describe the
local states of agents and do not explicitly describe time.

Schneider and Sidiropoulos define a notion of anonymity with respect to
a set $A$ of events.  Typically, $A$ consists of 
events 
of the form
$i.a$ for a fixed action $a$, where $i$ is an agent in some set that we
denote $I_A$. 
Intuively, anonymity with respect to $A$ means that if any event in $A$
occurs, it could equally well have been any other event in $A$.
In particular,
this means that if an agent in $I_A$ performs $a$, it could
equally well have been any other agent in $I_A$.
Formally, 
given a set $\Sigma$ of possible events and $A \subseteq \Sigma$, 
let $f_A$ be a function on traces
that, given a trace $\tau$, returns a trace $f_A(\tau)$ that is
identical to $\tau$ except that every event in $A$ is replaced by a
fixed event $\alpha \notin \Sigma$.  
A process $P$ is {\em strongly anonymous} on $A$ if  
$f_A^{-1}(f_A(P)) = P$,
where we identify
$P$ with its associated set of traces. This means that all the
events in $A$ are interchangeable; by 
replacing any event in $A$ with any other
we would still get a valid trace of $P$.

Schneider and Sidiropoulos give several very simple examples that are
useful for clarifying this definition of anonymity. One is a system
where there are two agents who can provide donations to a charity,
but where only one
of them will actually do so. Agent $0$, if she gives a donation,
 gives \$5, and agent $1$ gives \$10. This is followed by a ``thanks''
from the charity. The events of interest are ``0.gives'' and ``1.gives''
(representing events where $0$ and $1$ make a donation),
``\$5'' and ``\$10'' (representing the charity's receipt of the donation),
``thanks'', and ``STOP'' (to signify that the process
has ended). There are two possible traces:
\begin{enumerate}
\item 0.gives $\rightarrow$ \$5 $\rightarrow$ ``thanks'' $\rightarrow$
STOP.
\item 1.gives $\rightarrow$ \$10 $\rightarrow$ ``thanks'' $\rightarrow$
STOP.
\end{enumerate}
The donors require anonymity, and so we require that the CSP process is
strongly anonymous on the set \{0.gives, 1.gives\}. In fact, this
condition is not satisfied by the process, because ``0.gives'' and
``1.gives'' are not interchangeable. This is because ``0.gives'' must
be followed by ``\$5'', while ``1.gives'' must be followed by ``\$10''.
Intuitively, an agent who observes the traces can determine the
donor by looking at the amount of money donated.

We believe that Schneider and Sidiropoulos's definition is best
understood as trying to
capture the intuition that an observer who sees all the events generated
by $P$, except for events in $A$, does not know which event
in $A$ occurred. 
We can make this precise by translating Schneider and Sidiropoulos's
definition into our framework.
The first step is to associate with each process $P$ 
a corresponding set of runs $\R_P$.
We present one reasonable way of doing so here, which suffices for our
purposes. 
In future work, we hope to explore the connection between CSP and the runs and
systems framework in more detail.

Recall that a run is an infinite sequence
of global states of the form $(s_e, s_1, \ldots, s_n)$, where each $s_i$
is the local state of
agent $i$, and $s_e$ is the state of the environment.
Therefore, to
specify a set of runs, we need to describe the set of agents, and then
explain how to derive the local states of each agent for each run.
There is an obvious problem here: 
CSP has no analogue of agents and local states.
To get around this, we could simply tag all events
with an agent (as Schneider and Sidiropoulos in fact do for the events
in $A$).  
However, for our current purposes, a much simpler approach will do.
The only agent we care about is a (possibly mythical) observer who
is able to observe every event except the ones in $A$.  Moreover, for
events in $A$, the observer knows that something happened (although not
what).  There may be other agents in the system, but their local states
are irrelevant.  We formalize this as follows.

Fix a process $P$ over some set $\Sigma$ of events, and let $A \subseteq \Sigma$.  
Following Schneider and Sidiropoulos, for the purposes of this
discussion, assume that $A$ consists of events of the form $i.a$, where
$i \in I_A$ and $a$ is some specific action.
We say that a system $\R$ is {\em compatible with $P$\/} if there
exists some agent $o$ such that the following two conditions hold:
\begin{itemize}
\item for every run $r \in \R$ and every time $m$, there exists a trace
$\tau \in P$
such that $\tau = r_e(m)$ and $f_A(\tau) = r_o(m)$;
\item for every trace $\tau \in P$, there exists a run $r \in \R$ such
that $r_e( | \tau | ) = \tau$ and $r_o( | \tau | ) = f_A(\tau)$
(where $|\tau|$ is the number of events in $\tau$).
\end{itemize}
Intuitively, $\R$ represents $P$ if (1) for every trace $\tau$ in $P$,
there is a point $(r,m)$ in $\R$ such that, at this point, exactly the
events in $\tau$ have occurred (and are recorded in the environment's
state) and $o$ has observed $f_A(\tau)$, and (2) for every point $(r,m)$
in $\R$, there is a trace $\tau$ in $P$ such that precisely the events
in $r_e(m)$ have happened in $\tau$, and $o$ has observed $f_A(\tau)$ at
$(r,m)$.  
We say that the interpreted system $\I = (\R,\pi)$ is {\em compatible
with $P$\/} if $\R$ is compatible with $P$ and 
if $(\I,r,m) \models \theta(i,a)$ whenever the event 
$i.a$ is in the 
event sequence $r_e(m')$ for some $m'$.

We are now able to make a formal connection between our definition of
anonymity and that of Schneider and Sidiropoulos.
As in the setup of Proposition~\ref{pro:minimally}, we assume that an
anonymous action $a$ can be performed only once in a given run.

\thm\label{thm:stronganonthm}
If $\I = (\R,\pi)$ is compatible with $P$, then 
$P$ is strongly anonymous on the alphabet $A$ if and only if
for every agent
$i \in I_A$, the action $a$ performed by $i$ is anonymous up to $I_A$ with
respect to $o$ in $\I$.
\ethm

\prf
Suppose that $P$ is strongly anonymous on the alphabet $A$ and that 
$i \in I_A$.
Given a point $(r,m)$, suppose that $(\I,r,m) \models \theta(i,a)$, 
so that the event $i.a$ appears in $r_e(n)$ for some $n \geq m$.
We must show that
$(\I,r,m) \models P_o[\theta(i',a)]$ for every $i' \in I_A$,
that is, that $a$ is anonymous up to $I_A$ with respect to $o$.
For any $i' \in I_A$, this requires showing that there 
exists a point $(r',m')$ such that
$r_o(m) = r'_o(m')$, 
and $r'_o(n')$ includes $i'.a$, for some $n' \geq m'$. Because $\R$ 
is compatible with $P$, there exists $\tau \in P$ such that $\tau =
r_e(n)$ and $i.a$ appears in $\tau$. Let $\tau'$ be the trace identical to $\tau$
except that $i.a$ is replaced by $i'.a$. Because $P$ is strongly anonymous
on $A$, $P = f_A^{-1}(f_A(P))$, and $\tau' \in P$. By compatibility, there
exists a run $r'$ such that $r_e'(n) = \tau'$ and $r_o'(n) = f_A(\tau')$.
By construction, $f_A(\tau) = f_A(\tau')$, so $r_o(n) = r'_o(n)$. Because the
length-$m$ trace prefixes of $f_A(\tau)$ and $f_A(\tau')$ are the same,
it follows that $r_o(m) = r'_o(m)$.
Because $(\I,r',m) \models \theta(i',a)$, $(\I,r,m) \models P_o[\theta(i',a)]$
as required.

Conversely, suppose that for every agent $i \in I_A$, the action $a$
performed by $i$
is anonymous up to $I_A$ with respect to $o$ in $\I$. We must show that
$P$ is strongly anonymous.  
It is clear that $P \subseteq
f_A^{-1}(f_A(P))$, so we must show only that $P \supseteq 
f_A^{-1}(f_A(P))$.  So suppose that $\tau \in f_A^{-1}(f_A(P))$.  
If no event $i.a$ appears in $\tau$, for any $i \in I_A$, then $\tau \in P$
trivially. Otherwise, some $i.a.$ does appear. Because $\tau \in f_A^{-1}(f_A(P))$,
there exists a trace $\tau' \in P$ that is identical to 
$\tau$
except that $i'.a$
replaces $i.a$, for some other $i' \in I_A$. Because $\R$ is compatible with $P$,
there exists a run $r' \in R$ such that $r'_o(m) = f_A(\tau')$ and $r'_e(m) = \tau'$
(where $m = |\tau'|$). Clearly $(\I,r',m) \models \theta(i',a)$ so, by
anonymity, 
$(\I,r',m) \models P_o[\theta(i,a)]$, and there exists a run $r$ such that 
$r_o(m) = r'_o(m)$ and $(\I,r,m) \models \theta(i,a)$. Because the action $a$ can
be performed at most once, the trace $r_e(m)$ must be equal to $\tau$. By 
compatibility, $\tau \in P$ as required.
\eprf

Up to now, we have assumed that the observer $o$ has access to all the
information in the system except which event in $A$ was performed.  
Schneider and Sidiropoulos extend their definition of strong anonymity to
deal with agents that have somewhat less information.  They capture ``less
information'' using {\em abstraction operators}.
Given a process $P$, there are several
abstraction
operators that can give us a new process. For example the {\em hiding operator},
represented by $\backslash$, hides all events in some set $C$. That is,
the process $P \backslash C$ is the same as $P$ except that all events in
$C$ become internal events of the new process, and are not included in 
the traces associated with $P \backslash C$. 
Another abstraction operator, the renaming
operator,
has already appeared in the definition
of strong anonymity: 
for any set $C$ of events, we can consider the function $f_C$ that maps
events in $C$ to a fixed new event.  The difference between hiding and
renaming is that, if events in $C$ are hidden, the observer is not even
aware they took place.  If events in $C$ are renamed, then the observer
is aware that some event in $C$ took place, but does not know which one.

Abstraction operators such as these provide a useful way to model a
process or agent who has a distorted
or limited view of the system. In the context of
anonymity, they allow anonymity to hold with respect to an observer with a
limited view of the system in cases where it would not hold with respect to an
observer who can see everything. In the anonymous donations example,
hiding
the events \$5 and \$10, i.e., the amount of money donated, would make the
new process $P \backslash \{\$5, \$10\}$ strongly anonymous on the set of
donation events. Formally, given an abstraction operator $ABS_C$ on a set
of events $C$, we have
to check the requirement of strong anonymity on the process $ABS_C(P)$
rather than on the process $P$.

Abstraction is easily captured in our framework. It amounts simply to
changing the local state of the observer. For example, anonymity of 
the process $P \backslash C$ in our framework corresponds to anonymity of the
action $a$ for every agent in $I_A$ with respect to an observer whose
local state at the point $(r,m)$ is $f_A(r_e(m)) \backslash C$.  We omit the obvious
analogue of Theorem~\ref{thm:stronganonthm} here.

A major advantage of the runs and systems framework is that definitions of
high-level properties such as anonymity do not depend on the local states
of the agents in question. If we want to model the fact that an observer has
a limited view of the system, 
we need only modify her local state to reflect this fact. 
While some limited views are naturally captured by CSP abstraction
operators, others may not be.  The definition of anonymity should not
depend on the existence of
an appropriate abstraction operator 
able to capture the limitations of a particular observer.

As we have demonstrated, our approach to anonymity is compatible with the
approach
taken in~\cite{schneider96}. Our definitions are stated in terms of
actions, agents, and knowledge, and are thus
very intuitive and flexible. The generality of runs and systems allows us
to have simple definitions that apply to a wide variety of systems and
agents. The low-level CSP
definitions, on the other hand, are more operational than ours, and this
allows easier model-checking and verification. Furthermore, there are
many advantages to using process algebras in general: systems can often
be represented much more succinctly, and so on. This suggests that
both approaches have their advantages.
Because CSP
systems can be represented in the runs and systems framework, however,
it makes perfect sense to define anonymity for CSP processes using
the knowledge-based definitions we have presented here. If our
definitions turn out to be equivalent to more low-level CSP definitions,
this is ideal, because CSP model-checking programs can then be used for
verification. A system designer simply needs to take care that the
runs-based system derived from a CSP process (or set of processes)
represents the local states of the different agents appropriately.

\subsection{Anonymity and Function View Semantics}\label{sec:functionviews}
Hughes and Shmatikov \citeyear{hughes02} introduce {\em function views}
and function-view {\em opaqueness} as a way of
expressing a variety of information-hiding properties in a succinct and uniform way.
Their main insight is that requirements such as anonymity involve restrictions
on relationships between entities such as agents and actions. Because these relationships
can be expressed by functions from one set of entities to another, hiding information
from an observer amounts to limiting an observer's view of the function
in question. For example, anonymity properties are concerned with whether or not
an observer is able to connect actions with the agents who performed them. By
considering the function from the set of actions to the set of agents who performed
those actions, and specifying the degree to which that function must be opaque to 
observers, we can express anonymity using the 
function-view approach.

To model the uncertainty associated with a given function, Hughes and
Shmatikov define a notion of
{\em function knowledge} to explicitly represent an observer's 
partial knowledge of a function.
Function knowledge focuses on three particular aspects of a function:
its graph, image, and kernel. 
(Recall that the {\em kernel\/} of a function $f$ with domain $X$ is the
equivalence relation {\em ker\/} on $X$ defined by $(x,x') \in
\mbox{{\it ker}}$ iff $f(x) = f(x')$.)
{\em Function knowledge\/} of type $X \rightarrow Y$ is a triple
$N = (F,I,K)$, where $F \subseteq X \times Y$, $I \subseteq Y$, and $K$
is an equivalence relation on $X$.  A triple $(F,I,K)$ is {\em
consistent with $f$\/} if $f \subseteq F$, $I \subseteq \im f$, and
$K \subseteq \kernel f$. Intuitively, a triple $(F,I,K)$ 
that is consistent with $f$ represents
what an agent might know about
the function $f$. Complete knowledge of a function $f$, for example, would be 
represented by the triple $( f, \im f, \kernel f )$.

For anonymity, and for information hiding in general,
we are interested not in what an agent
knows, but in what an agent does {\em not\/} know.
This is formalized 
by Hughes and Shmatikov
in terms of
opaqueness conditions for function knowledge. 
If $N = \langle F, I, K \rangle$ 
is consistent with $f: X \rightarrow Y$, then, for example,
$N$ is {\em $k$-value opaque\/} if $|F(x)| \geq k$  for all $x \in X$.
That is, $N$ is $k$-value opaque if there are $k$ possible candidates
for the value of $f(x)$, for all $x \in X$. Similarly, $N$ is
{\em $Z$-value opaque\/} if $Z \subseteq F(x)$
for all $x \in X$. In other words, for each $x$ in the domain of $f$,
no element of $Z$ can be ruled out as a candidate for $f(x)$. 
Finally, $N$ is {\em absolutely value opaque\/} if that $N$ is $Y$-value opaque.

Opaqueness conditions are closely related to the 
nonprobabilistic definitions of anonymity given in Section~\ref{sec:anonymity}.
Consider functions from $X$ to $Y$, where $X$ is a set of actions
and $Y$ is a set of agents, and suppose that some function $f$ is the function
that, given some action, names the agent who performed the action. If we have 
$k$-value opaqueness for some view of $f$ (corresponding to some
observer $o$), this means, essentially, that
each action $a$ in $X$ is $k$-anonymous with respect to $o$.
Similarly, the view is $I_A$-value
opaque if the action is 
anonymous up to $I_A$ for each agent $i \in I_A$. Thus, function
view opaqueness provides a concise way of describing anonymity properties,
and information-hiding properties in general.

To make these connections precise, we need to explain how function views
can be embedded within the runs and systems framework. 
Hughes and Shmatikov already show how we can define function
views using
Kripke structures, the standard approach for giving semantics
to knowledge.  A minor modification of their approach works in systems
too.  Assume we are interested in who performs an action $a \in X$, where
$X$, intuitively, is a set of ``anonymous actions''.  Let $Y$ be the set of agents, 
including a ``nobody agent'' denoted $N$, and let $f$ be a 
function from $X$ to $Y$.  Intuitively,
$f(a) = i$ if agent $i$ 
performs action $a$,
and $f(a) = N$ if no agent performs action $a$.
The value of the
function $f$ will depend on the point.  Let $f_{r,m}$ be the value of
$f$ at the point $(r,m)$.  Thus, $f_{r,m}(a) = i$ if
$i$ performs $a$ in run $r$.
\footnote{Note that for $f_{(r,m)}$ to be well-defined, it must
be the case that only one agent can ever perform a single action.
}
We can now easily talk about function opaqueness with respect to an
observer $o$. For example, $f$ is $Z$-value opaque
at the point $(r,m)$ with respect to $o$ if, for 
all $z \in Z$, there exists a point $(r',m')$ such that $r'_o(m') = r_o(m)$
and $f_{(r',m')}(x) = z$. In terms of knowledge, $Z$-value opaqueness
says that for any value $x$ in the range of $f$, 
$o$ thinks it possible that
any value $z \in Z$ could be the result of $f(x)$. Indeed, Hughes and
Shmatikov say that function-view opaqueness, defined in terms of Kripke
structure semantics, is closely related to epistemic logic. The following
proposition makes this precise; it would be easy to state similar propositions
for other kinds of function-view opaqueness. 

\pro\label{pro:zvaluethm}
Let $\I = (\R, \pi)$ be an interpreted system that satisfies
$(\I,r,m) \models f(x) = y$ whenever $f_{(r,m)}(x) = y$.
In system $\I$, $f$ is $Z$-value opaque for observer $o$ at the point $(r,m)$ if
and only if
\[ (\I,r,m) \models \bigwedge_{x \in X} \bigwedge_{z \in Z} P_o[f(x) = z]. \]
\epro

\prf
This result follows immediately from the definitions.
\eprf

Stated in terms of knowledge, function-view opaqueness already looks 
a lot like our definitions of anonymity. 
Given $f$ (or, more precisely, the set $\{f_{(r,m)}\}$ of functions) mapping
actions to agents, 
we can state a theorem connecting anonymity to function-view opaqueness.
There are two minor issues to deal with, though. First, our definitions
of anonymity are stated with respect to a single action $a$, while the 
function $f$ deals with a {\em set} of actions. We can deal with this
by taking the domain of  $f$ to be
the singleton $\{a\}$. Second, our definition
of anonymity up to a set $I_A$ requires the observer to suspect
agents in $I_A$ only
if $i$ actually performs the action $a$.
(Recall this is also true for Syverson and Stubblebine's definitions.)
$I_A$-value opaqueness requires the observer to think many agents
could have performed an action even if nobody has. 
To deal with this, we require opaqueness only
when the action has been performed
by one of the agents in $I_A$.
\commentout{
Last, opaqueness provides anonymity for {\em all} agents who might
want to perform $a$. To restrict our attention to agents in $I_A$,
we require that $f_{(r,m)}(a) \in I_A$ for all points $(r,m) \in \P(\R)$.
}
\commentout{
If $f$ maps from an action $a$ to the agent who performed
$a$, $f_{(r,m)}(a)$ may be undefined if nobody has actually performed
$a$ by time $m$ in the run $r$. For functions that might possibly
be undefined, we can say that $f$ is $Z$-value opaque if 
for all points $(r,m) \in \P(\R)$, for all $x \in X$ such that 
$f_{(r,m)}(x)$ is well defined, and for 
all $z \in Z$, there exists a point $(r',m')$ such that $r'_o(m') = r_o(m)$
and $f_{(r',m')}(x) = z$.
}

\thm\label{thm:functionviews}
Suppose that $(\I,r,m) \models \theta(i,a)$ exactly if $f_{(r,m)}(a) = i$.
Then action $a$ is anonymous up to $I_A$ with respect to $o$
for each agent $i \in I_A$ if and only if at all points $(r,m)$ such
that $f_{(r,m)}(a) \in I_A$,
$f$ is $I_A$-value opaque with respect to $o$.
\ethm

\prf
Suppose that $f$ is $I_A$-value opaque, and let $i \in I_A$ be given.
If $(\I,r,m) \models \theta(i,a)$, then $f_{(r,m)}(a) = i$. 
We must show that, for all $i' \in I_A$, 
$(\I,r,m) \models
P_o[\theta(i',a)]$. 
Because $f$ is $I_A$-value opaque at $(r,m)$,
there exists a point 
$(r',m')$ such that $r'_o(m') = r_o(m)$ and $f_{(r',m')}(a) = i'$.
Because $(\I,r',m') \models \theta(i',a)$, $(\I,r,m) \models P_o[\theta(i',a)]$.

Conversely, suppose that for each agent $i \in I_A$, 
$a$ is anonymous up to $I_A$ with respect to $o$.
Let $(r,m)$ be given such that $f_{(r,m)}(a) \in I_A$,
and let that $i = f_{(r,m)}(a)$. 
It follows that $(\I,r,m) \models \theta(i,a)$.
For any $i' \in I_A$, $(\I,r,m) \models P_o[\theta(i',a)]$, by
anonymity. Thus there exists a point $(r',m')$ such that 
$r'_o(m') = r_o(m)$ and $(\I,r',m') \models \theta(i',a)$.
It follows that $f_{(r',m')}(a) = i'$, and that $f$ is
$I_A$-value opaque.
\eprf

As with Proposition~\ref{pro:zvaluethm}, it would be easy to state
analogous theorems connecting our other definitions of anonymity,
including minimal anonymity, total anonymity, and $k$-anonymity,
to other forms of function-view opaqueness. We omit the details here.

The assumptions needed to prove Theorem~\ref{thm:functionviews} illustrate
two ways in which our approach may seem to be less general than the
function-view  
approach. First, all our definitions are given with respect to a single
action, rather than with respect to a set of actions. However, it is 
perfectly reasonable to specify that all actions in some set $\A$ of actions 
be anonymous. Then we could modify Theorem~\ref{thm:functionviews}
so that the function $f$ is defined on all actions in $\A$. 
(We omit the details.) Second, our definitions of anonymity only
restrict the observer's  
knowledge if somebody actually performs the action. This is simply a different
way of defining anonymity. As mentioned previously,
we are not trying to give a definitive definition of anonymity, and it certainly
seems reasonable that someone might want to define or specify anonymity using the
stronger condition. At any rate, it would be straightforward to modify our 
definitions so that the implications, involving $\theta(i,a)$, are not included.

Hughes and Shmatikov argue that epistemic logic is a useful language 
for expressing anonymity specifications, while CSP is a useful language
for describing and specifying systems.  We agree with both of
these claims.  They propose function views as a useful interface to
mediate between the two.  We have tried to argue here that no mediation
is necessary, since the multiagent systems framework can also be used for
describing systems. (Indeed, the traces of CSP can essentially be viewed as runs.)  
Nevertheless, we do believe that function views can be the
basis of a useful language for reasoning about some aspects of
information hiding. We can well imagine
adding abbreviations to the language that let us talk directly about
function views.  (We remark that we view these abbreviations as
syntactic sugar, since these are notions that can already be expressed
directly in terms of the knowledge operators we have introduced.)  

On the other hand, we believe that function views are not expressive
enough to capture all aspects of information hiding.  One obvious
problem is adding probability.  While it is easy to
add probability to systems, as we have shown, and to 
capture interesting probabilistic notions
of anonymity, it is far from clear how to do this if we take function
views triples as primitive.   

To sum up, we would argue that to reason about
knowledge and probability, we need to have possible worlds as the
underlying semantic framework.  Using the multiagent systems approach
gives us possible worlds in a way that makes it particularly easy to
relate them to systems.  Within this semantic framework, function views
may provide a useful syntactic construct with which to reason about
information hiding.

\section{Discussion}\label{sec:conclusion}

We have described a framework for reasoning about information hiding
in multiagent systems, and have given general definitions of anonymity
for agents acting in such systems. We have also compared and contrasted 
our definitions to other similar definitions of anonymity. Our knowledge-based
system framework provides a number of advantages:
\ul
\item We are able to state information-hiding properties succinctly
and intuitively, and in terms of the knowledge of the observers or
attackers who interact with the system.
\item Our system has a well-defined semantics that lets us reason about
knowledge in
systems of interest, such as systems specified using process algebras 
or strand spaces.
\item We are able to give straightforward probabilistic definitions of 
anonymity, and of other related information-hiding properties.
\eul

There are a number of issues that this paper has not addressed.
We have focused almost exclusively on properties of anonymity, 
and have not considered related notions, such as \emph{pseudonymity} and
\emph{unlinkability}~\cite{hughes02,terminology}. 
There seems to be no intrinsic difficulty capturing these notions in our
framework.  For example, one form of message  unlinkability
specifies that no two messages sent by an
anonymous sender can be ``linked'', in the sense that an observer can
determine that both 
messages were sent by the same sender. More formally, two actions $a$
and $a'$  
are linked with respect to an observer $o$ if $o$ knows
that there 
exists an agent $i$ who performed both $a$ and $a'$. This definition 
can be directly captured using knowledge.  Its negation says 
that $o$ considers it possible 
that there exist two distinct agents who performed $a$ and $a'$; 
this can be viewed as a definition of \emph{minimal unlinkability}.
This minimal requirement can be strengthened, exactly as our definitions of
anonymity were, 
to include larger numbers of distinct agents, probability, and so on. 
Although we have not worked out the details, we believe  that our
approach will be similarly applicable to
other definitions of information hiding.

Another obviously important issue is
checking whether 
a given system specifies
the knowledge-based properties we have introduced. 
The standard technique for doing this is model checking.
Recent work on the problem of model checking in
the multiagent systems framework suggests that this may be viable.
Van der Meyden~\citeyear{vandermeyden98} discusses 
algorithms and complexity results for model checking a wide range of
epistemic formulas in the runs and systems framework, and van der Meyden and 
Su~\citeyear{vandermeyden02} use these results to verify the 
dining cryptographers protocol~\cite{Chaum88}, using formulas much
like those described in 
Section~\ref{sec:diningexample}.
Even though model checking of formulas involving knowledge seems to
be intractable for large problems, these results are a promising first
step towards being able to use knowledge for both the specification and
verification of anonymity properties. 
Shmatikov~\citeyear{shmat02}, for example, analyzes the Crowds system
using the 
probabilistic model checker PRISM~\cite{KNP01}.  This is a
particularly good example of how definitions of 
anonymity can be made precise using logic and probability, and how
model-checking 
can generate new insights into the functioning of a deployed protocol.

Finally, it is important to note that the examples considered in this
paper do 
not reflect the state of the art for computational anonymity. Anonymity
protocols 
based on DC-nets, while theoretically interesting, have not been widely
deployed; in practice, protocols based on mixes and message-rerouting 
are much more common.
We used the dining cryptographer's problem as a running
example here mainly 
because of its simplicity, but it remains to be seen whether our general
approach will be as 
illuminating for more complicated protocols. There are reasons to
believe that it will be.  Shmatikov's analysis of Crowds shows
that a logic-based approach can be useful  
for analyzing protocols based on message-rerouting. Furthermore, we
believe that formalizing 
anonymity protocols using techniques like ours is worthwhile even if
formal 
verification is impractical or impossible. 
It forces system designers to 
think carefully about information-hiding requirements, which can often
be tricky,  
and provides a system-independent framework for
comparing the anonymity  
guarantees provided by different systems. 

We described one way to generate a set of runs from a CSP process $P$, 
basically by recording all the events in the state of the environment
and describing some observer $o$ who is able to observe a subset of
the events. This translation was useful for comparing our abstract
definitions of anonymity to more operational CSP-based definitions.
In future work we hope to further explore the connections between the
runs and systems framework and tools such as CCS, CSP, and the spi 
calculus~\cite{abadi97}. 
In particular, we are interested in the 
recent work of Fournet and Abadi~\citeyear{fournet02}, who use the 
applied pi calculus to model {\em private authentication}, according to which
a principal in a network is able to authenticate herself to
another principal while remaining anonymous to other ``nearby''
principals.
A great deal of work in computer security has formalized
information-hiding properties using these tools. Such work often reasons
about the knowledge of various agents in an informal way, and then
tries to capture knowledge-based security properties
using one of these formalisms.
By describing canonical translations from these formalisms to the runs
and systems framework, we hope to be able to demonstrate formally
how such definitions of security do (or do not) capture notions of knowledge.

\paragraph{Acknowledgments:}  We thank Dan Grossman, Dominic Hughes, 
Vitaly Shmatikov, Paul Syverson, Vicky Weissman, 
the CSFW reviewers
(who were anonymous 
up to $P_{CSFW}$, the program committee), 
and the JCS referees,
for useful discussions,
helpful criticism, and 
careful copy-editing.

\bibliographystyle{chicago}
\bibliography{joe,z}

\end{document}